\documentclass[11pt]{asaproc}
\usepackage[dvips]{graphicx}             
\usepackage{pslatex}	
\usepackage{amsmath,amssymb,amsthm}             
\usepackage{eqnarray}
\usepackage{fancyhdr}                    
\usepackage{subfig}                   
\usepackage[vcentering,dvips]{geometry}  
\usepackage{rotating}
\usepackage{enumitem}                    
\usepackage[ruled,vlined]{algorithm2e}      
\usepackage{float}                         
\usepackage[authoryear ]{natbib} 
\usepackage{xcolor}
\usepackage{setspace}
\usepackage[utf8]{inputenc}
 \usepackage{varwidth}
\usepackage[utf8]{inputenc}
\usepackage{longtable}
\usepackage{booktabs}
\usepackage{multirow,booktabs}
\usepackage{times}

\title{ Monotonic Nonparametric Dose Response Model}

\author{Faten S. Alamri\thanks{ Princess Nourah Bint Abdulrahman University,Riyadh ,Saudi Arabia, PO Box 84428 and Virginia Commonwealth University, Richmond, VA 23284, USA} \and   Edward L. Boone\thanks{ Virginia Commonwealth University, Richmond, VA 23284, USA} \and  David J. Edwards\thanks{ Virginia Commonwealth University, Richmond, VA 23284, USA} }
\begin{document}

	\maketitle

\begin{abstract}
Toxicologists are often concerned with determining the dosage to which an individual can be exposed with an acceptable risk of adverse effect. These types of studies have been conducted widely in the past, and many novel approaches have been developed. Parametric techniques utilizing ANOVA and non-linear regression models are well represented in the literature. The biggest drawback of parametric approaches is the need to specify the correct model. Recently, there has been an interest in non-parametric approaches to tolerable dosage estimation. In this work, we focus on the monotonically decreasing dose response model where the response is a percent to control. This poses two constraints to the non-parametric approach. The dose-response function must be one at control (dose=0), and the function must always be positive. Here we propose a Bayesian solution to this problem using a novel class of non-parametric models. A basis function developed in this research is the Alamri Monotonic spline (AM-spline). Our approach is illustrated using two simulated data and two experimental dataset from pesticide related research at the US Environmental Protection Agency.
\end{abstract}


\section{Introduction}
Evaluating the risk of exposure to chemicals starts by measuring the side effects of chemical on an experiment specimen, it is a critical step in pharmaceutical drug development and in other chemical areas as well. Many toxicology studies are performed on rodents and in some cases require sacrificing the rodent to get the endpoints measurement that indicates the level of chemical side effects. Toxicologists are searching to develop methods that determine the dose which corresponds to the research targeted dose-response effect. Are the chemical dangers at any dose or is there is a specific limit of a safe dose? How much of the chemical can we consume without getting into the bad side effect? These question motivate our research in developing new dose response model that answer these question. Toxicologist goal is to find the safe dose with an acceptable side effect. The literature introduced several methods that deal with dose-response determination. The Benchmark dose (BMD) is a method to find the maximum tolerable dose that produces a prespecified acceptable side effect level on the experimental specimen. Other methods, the No Observable Adverse Effect Level (NOAEL) which means the highest dose at which there was no observed effect and the Low Observed Adverse Effect Level (LOAEL) approach \citet{crump1984new} and \citet{hunt2008summary}. There are many other dose-response model threshold \citet{crump1984new}. Which determine the safest dosage, the dangerous dosage or any other inquiry. The method we consider in our research is the Effective Dose $ED_{\gamma}$, where $\gamma$ is the level of side effects on the experiment specimen which could be $50\%, 75\%, 90\% $ or any other number of interest. $ED_{50}$ specifically is the BMD type that we will use in this article, which is the dose that causes a 50\% reduction in the average response.
BMD is a favorable method than NOAEL since it presumes the dose-response model shape that depends on the data more. For more about benchmark dose estimation, see \citet{shao2018web}.

Most statistical methodology for dose-response studies has been using the parametric models such as in \citet{holland2015optimal}, where they used the Log-normal function, Log-logistic function; Exponential function, Gaussian function; Logistic function, Gompertz  function and the Weibull function. In parametric model all the parameters are in finite-dimensional parameter spaces. Fitting of the parametric dose-response model requires knowledge about the parameters of the model. More about parametric dose response model in \citet{pinheiro2014model}, \citet{hunt2004parametric} and \citet{hunt2008summary}. The parametric model are flexible in fitting data when information about the parameters,  data shape and type are provided. That makes the parametric model struggle in fitting model to the unknown parameters or unexpected data shape. This where nonparametric models becomes valuable in fitting dose-response data where the parameters are unknown and the shape of the data is unpredictable.

The difficulty that faced the parametric model motivates us to develop a nonparametric model that is more flexible in fitting unknown parameters and rough $"wiggly"$ data.  Nonparametric regression is different than the parametric regression by that its capture unexpected feature of the data uses a different shape of the functional relationship. The Nonparametric regression is a type of regression analysis that known as a distribution-free with models that are infinite-dimensional, as in the following format
$$y_{i}= m(x_{i})+\epsilon_{i},\quad i=1,....,n $$
Where $0 \leq x_{1} \leq x_{2} \leq ....\leq x_{N}\leq 1 $ and the $\epsilon_{i}$ are independent draws from normal$(0,\sigma^2)$ with unknown $\sigma >0 $ and m(.) is the unknown smooth and flexible function. 
Different types of nonparametric modelling are introduced in \citet{wandbook}, \citet{thomas1983nonparametric}, \citet{corder2014nonparametric} and \citet{gibbons2011nonparametric}, such as, regression splines, smoothing splines, kernel methods including local regression, series-based smothers, and wavelets.

We are proposing a monotonic nonparametric model, which is flexible to fit the dose-response data. Different non-parametric methods introduced to estimate the monotonic dose-response curve using the bootstrap as \citet{dilleen2003non}. \citet{delecroix1996functional} used the kernel method to estimate the monotonic dose-response curve under general shape restriction. Out of all the nonparamteric model types we consider the spline model. So we are developing a monotonic nonparamtric model for a monotonic decreasing data.
Spline is a nonparametric regression technique that is written as a combination of basis function. It has a basis function representation which fits a smooth curve between points in the data called the knots that is in the interval [L, U] with specific constraints. All splines follows this model $$y_{i}=f(x_{i})+\epsilon_{i}, f_{i}\in [0,1] ,  i=1,2,.....,n $$
where $f(x_{i})$ is  spline basis function and $\epsilon_{i}$ is the white noise \citet{wegman1983splines}.
There are many types of spline such as M, B, P, cubic, linear and quadratic spline for more about spline types read \citet{semiparam}. Every spline have $t$ which are the set of basis functions that connected linearly which differ in each spline. This article will contribute to a new monotonic spline. Several monotonic splines are in the literature but our proposed spline is different as it has specific constraints and it has a more general structure. I-splines is a monotone spline constructed by a non-negative linear combination coefficient as \citet{Ramsy1988} who used the integrated B-spline as basis function to maintain monotonicity. \citet{Ramsy} fitted the constrained curve using non-Bayesian methods. \citet{he1998monotone} mentioned that  I-spline faces uncertainty when fitting it to real data. So,they proposed monotonic spline using a quadratic spline as the basis function. An isotonic spline is a monotonic spline that depends on the cubic spline, and it’s non-decreasing on a specific integral by certain constraints as mentioned in \citet{wang2008isotonic}. 
\citet{xue2010distribution} moved to higher than the quadratic spline order using \citet{he1998monotone} methods, which is computationally longer but they made monotonicity possible for any penalized splines (PS) order. Each spline has its own knots, these knots are a sequence of points that divides the spline interval to subintervals and its different in values and locations at different spline. It is known in the literature that the spline smoothness is controlled by the number of knots and their locations, thus number of knots are less than the data points. Knots have different selection methods; \citet{wold1974spline} states some recommendation for knots selection, his recommendation are upon the assumption of the cubic spline which needs modifications for a spline with degree greater than three. Cubic spline frequently used since no lower degree spline can interpolate through data endpoints that have exact derivative at each point\citet{wolberg1999monotonic}.   
Knots have properties we introduce some here: The knots are located on the data points. The Minimum of 4 to 5 observation should be between knots. Knots should not have more than one extrema, and one inflection point, both should be between knots. The extrema should be centred in the interval and inflection points should be close to knots points.

A general introduction of the theory in interpolating and smoothing splines is given by \citet{Wasserman:2006:NS:1202956} and \citet{green}. \citet{dimatteo2001bayesian} used fully Bayesian method for curve fitting with free-knot splines using the reversible-jump Markov chain Monte Carlo as a posterior sampling tool. Where the literature was studying the constraints on the parameters of B-splines, \citet{wood1994monotonic} used piecewise polynomial properties of a spline with conditions on monotonicity. That considered the cubic piecewise from \citet{hyman1983accurate} and extends the work to the cross-validation and confidence interval techniques.

Spline basis function known as 
\begin{equation} \label{eq:3}
f(x)=\sum ^{k}_{i=1} a_{i} x_{i}    , i=1,.....,n  
\end{equation}
Where $a_{i}$ set of non negative weights sums up to one and $x_{i}$ are the spline basis function over the knots $K_{i}$. Smoothing spline  is a popular technique with spline usage in \citet{silverman1985some} that provide a review of all possible smoothing methods. We used a smoothing parameter to control the smoothing fit of our proposed spline. Selecting the parameter something $\lambda$ or the knots $k$ is crucial and differs in each spline. Different selection method are proposed in the literature as \citet{xue2010distribution} used AIC criteria to select the number of interior knots $K_{n}$. 
Fitting spline using cross-validation was covered by \citet{wahba1975completely}, a valid mean square error method used to determine the correct degree of smoothing to a discrete data, and used Markov chain Monte Carlo (MCMC) which is the Bayesian sampling method to estimate the true smooth function and its derivative.

The test of monotonic regression function based on the critical bandwidth and the smoothing level imposes the non-parametric estimate to be monotonic in \citet{bowman1998testing}. Curves of a dose response model estimated by contracting a combination of smoothing spline and the non-negative properties of cubic B-spline was used as in \citet{kong2006monotone}. Our approach is similar to \citet{clyde2007nonparametric} where they used a regression function model as a linear combination of kernels. However, they used the general L\'evy processes as the prior distribution on the measure where we used the stick breaking prior distribution. Their approach also different than ours, since they are not considering the monotone regression in their model constraint, along with the utilization of a different processes. Another approaches that is similar is \citet{bornkamp2009bayesiangermen}, but they used a monotone increasing function and the Two-Sided Power distribution (TSP) as the spline bases function. They concluded that TSP is 10 to 15 times faster than using the Beta distribution function, more on TSP is found in\citet{van2002standard}. 
Many authors have contributed to this area in the past. Smooth monotone functions and the properties are introduced by \citet{ramsay1989binomial} such as the I-splines. I-spline is a monotone spline constructed by a non-negative linear combination coefficient constricted by \citet{Ramsy} who used the integrated B-spline as basis function to maintain monotonicity. Users of non-parametric smoothing techniques should utilize their judgement in deciding the estimated regression curve and the smoothing level.  The noise level which controls the smoothness of the curve is a subjective decision as \citet{hardle1990applied} used a software result to subjectively determine the smoothing level. Comparison of different non-parametric methods introduced in \citet{bhattacharyanonparametric}.

We consider the Bayesian framework using a normal likelihood and stick breaking prior to estimate the posterior predictive by the MCMC sampling method. In this paper, we proposed a new non-parametric model as an alternative to the parametric model for cases where the parametric models do not fit the data. In this research, a new spline model (AM-Spline) was developed, which matches the pathological example well. The AM-spline was used as a dose-response model which we develop an algorithm to define the tolerable region that contains the safest chemical dosage with an acceptable side effects. Several researchers have discovered other approaches in isotonic regression with smoothing consideration. For instance, \citet{wright1980isotonic}, \citet{mammen1991estimating}, \citet{hardle1990applied}, \citet{friedman1984monotone}, and \citet{kim2018nonparametric} who used the hierarchical Bayes framework, and characterization of stick-breaking process that allows unconstrained estimation of the monotone function. What motivated us is that all parametric and non-parametric models struggle in fitting a sinuous monotonic decreasing data.

Method for non-parametric monotone based on Bayesian analysis using isotonic regression were developed by \citet{neelon2004bayesian}, who defined the flat region of the dose-response curve constructed using a piecewise linear model with restricted prior distribution, along with the latent Markov process formulation that was used to simplify the computation and to form a smooth regression line. Another approach is the Semiparametric, which was suggested for dose-response analysis \citet{wheeler2012monotonic}. A semiparametric estimation has a constrained shape, which was introduced by \citet{wu2018semiparametric} for elasticity. They used penalized splines when applying the shape constraint on the fitted model, and their work was inspired by \citet{Ramsy}. Assumption of monotonicity on the dose-response curve, and continuous observation to define the confidence bands for isotonic dose-response curves found by \citet{korn1982confidence} is modified by \citet{lee1996estimation} while keeping monotonicity. Monotonic increasing function in the Bayesian framework considers a mixture of triangular distribution with unspecified dimension during the analysis the method is introduced by \citet{perron}. Their approach is not restricted to Bayesian, but could have other applications in the frequentest perspective. 

AM-spline is a spline model that accommodates and fits the unexpected rough monotonically decreasing data. Our new novel spline works with any possible statistical distributions, that makes it more adaptable to various data. In other words, it contributed to the dose-response model as we introduce here.

This paper is organized as follows: Section \ref{ME}, motivating example
Section \ref{Baysian}, contains the prior distribution, posterior computation and BMD/BMDL. 
We applied our method to possible application in Section \ref{Result}, while conclusions and discussion are offered in Section \ref{Discussion}.

\subsection{Motivating Example}\label{ME}

To motivate our approach, we reanalyzed the study of the Organophosphate Pesticide (OP) data by \citet{moser2005neurotoxicological}. We will show that the AM-spline model fits the data adequately and will compare it to other methods.  
A neurotoxic study was conducted using 349 rats to investigate the effects of OP, which is a common active pesticide in agriculture. OP data represent the side effects on the nervous system of the rats that have been measured and called the endpoints. \cite{moser2005neurotoxicological} tested multiple neurotoxicity endpoints in which we only consider the Blood cholinesterase (BloodCHE).

This endpoint was measured at the effect of two different pesticides Acephate (ACE) and Diazinon (DIA) with different level of doses. ACE dose level from 0 to 120 (mg/kg) and DIA dose level from 0 to 250 (mg/kg).

\begin{figure}[H] 
	\begin{center}
		\begin{tabular}{ c c }
			(a) & (b) \\
			\includegraphics[width=3in]{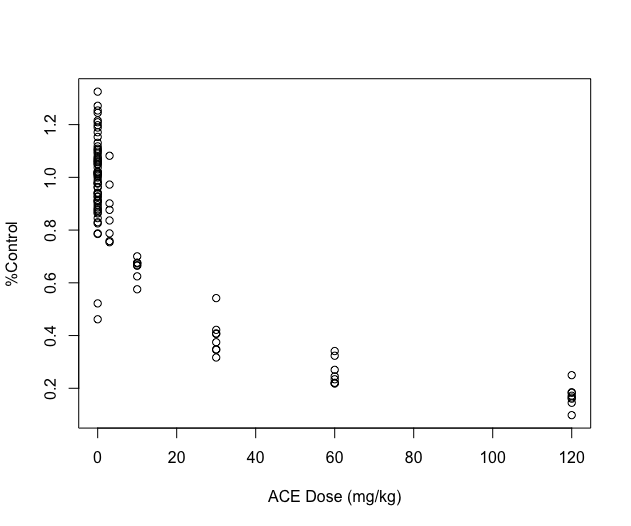} &
			\includegraphics[width=3in]{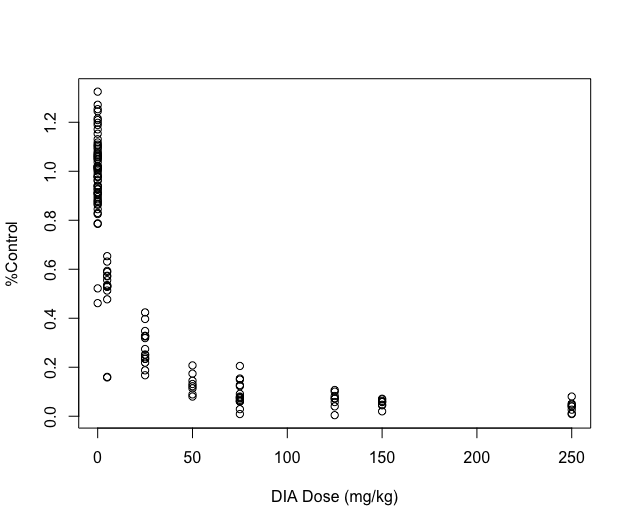} 
		\end{tabular}
	\end{center}\caption{ OP Data Blood considering the Blood Cholinesterase (BloodCHE) endpoint. In panel (a) is the ACE pesticide data and in panel (b) is DIA pesticide data, both data are monotonically decreasing. The x-axis is the chemical dosage (mg/kg) and the y-axis is the percent to control which is the percent to corresponding controls required to consider the variance of these controls.} \label{fig:OP data}%
\end{figure}

Each pesticide was consumed by the rat and absorbed into their system while the other pesticide had a dose level of zero. The continuous endpoint BloodCHE is measured and recorded from the rats system. 
In our research we considered each pesticide dosage individually and BloodCHE endpoint measurement which is available for each rat. See \citet{moser2005neurotoxicological} for more details about the data collection and measurement process.
Figure \ref{fig:OP data} shows that the OP data is monotonically decreasing. In panel (a) the OP data is considering the ACE pesticide. We see the data is monotonically decreasing with variation at zero dosage and the remaining data are gradually decreasing, most of the data are in the range of $0$ to $0.7$ percent to control. Panel (b) is the DIA pesticide data which is dramatically decreasing with some variation at zero dosage; most of the data fall in the range of $0$ to $0.7$ percent to control. The Figure shows that DIA data is denser then ACE data. The x-axis is the chemical dosage and the y-axis is the percent to control which is the percent to corresponding controls required to consider the variance of these controls. More about percent to control in \citet{feuerstein1997express}.

Fitting a model allows us to identify the region of the safe dose level. Understanding this region could help to address environmental public health questions, such as what is the safest dose of pesticide? The literature describes many dose-response models that fit the type of data we have but, to our best knowledge, none have used our proposed model with the specific constraints we have and none have used it in dose-response perspective.

\section{Methodology}\label{Methodology}

\subsection{Isotonic Regression} \label{IR} 
Isotonic Regression is an ordered set for the real numbers $\{y_1,y_2,.....,y_n\}$. The difficulty is finding $\{m_1,m_2,...,m_n\}$ that minimizes the $\sum_{n}^{1}(y_i-m_i)^2$ given the restriction $m_1\leq m_2 \leq ......\leq m_n $. A different approach to find the solution to that problem has been discovered by \citet{barlow1972isotonic}, yet their algorithm is complicated, with the basic idea introduced by \citet{friedman1984monotone}. It depends on choosing the scatter plot points starting with $y_i$ then moving to the right and stop at first place where $y_{i} > y_{i+1}$. When $y_{i+1}$ does not meet the monotone assumption, then $y_i$ and $ y_{i+1}$ will be replaced by their average $$\bar{y}= (y_i + y_{i+1})/2.$$ After finding the average they move to the left to make sure $y_{i-1}\leq \bar{y}$ if not, they pulled $y_{i-1}$ with $\bar{y}_{i}$ and $\bar{y}_{i-1}$ replacing the three with their average. They continued to the left until monotone assumption is satisfied, then get back to the right. This process of pulling the first violated point, and using the average of the two points is continued until reaching the right edge in \citet{friedman1984monotone} the solution to the dual problem. The goal for that monotone process was a monotone increasing process. Our goal is a monotone decrease so the difference will be that this monotone assumption $y_{i} > y_{i+1}$, and ours is $y_{i} < y_{i+1}.$

A generalized monotonic model fitted to the Bayesian analysis using isotonic regression model that was introduced by \citet{holmes2003generalized}. In order to make inference on posterior parameters of interest they adopt MCMC to sample the space of unconstrained models by varying the numbers and the location of points. Isotonic and antotonic are two different monotonically cases in nonparametric, so we move onto the next subsection to the general monotonic regression.

\subsection{Monotone Regression} \label{MR}
A monotonic function is a function where it is either entirely increasing or decreasing. 
The model for continuous and homoscedastic data 
\begin{equation} \label{eq:1}
y_{i}=\mu(x_{i}) + \epsilon_{i}   , i=1,.....,n  
\end{equation}
where $\epsilon_{i} \sim N(0,\sigma^{2})$ and $\mu(.)$ is a continuous monotonic. Variables $x_{i}$ have to be bounded between 0 and 1. Our continuous monotonic decreasing function is from zero to one. 
The $\mu^{0}(.)$ here have the following formula  
\begin{equation} \label{eq:2}
\mu(x)=\beta_{0} + \beta_{1}\mu^{0}(x) 
\end{equation}
Where $\mu^{0}(.)$ the probability distribution function of the continuous bounded variable between $0$ and $1$. $\beta_{0}$ is the intercept that represent the response at $0$ and on the other hand, $\beta{0}+\beta{1}$ represent the response at $1$. Many applications for $\beta_{0}$ and $\beta_{1}$ which have a clear cut interpretation, just as in the dose response model $\beta_{0}$ represent an inactive drug and $\beta_{1}$ represent the maximum effect of the drug.

The probability distribution function $\mu^{0}(.)$ prior is a discrete mixture of parametric distribution function $F(x,\zeta)$ of bounded continuous variables on interval [0,1], with parameters $\zeta \in \Xi $. So, the model is 
$$ \mu^{0}(x)= \int_{\Xi} F(x,\zeta)P(d\zeta)$$ 
where $P$ is a discrete mixing distribution on $\Xi$ \citet{bornkamp2009bayesiangermen}. The most used random probability measure is the Dirichlet Process because it has reasonable analytical properties for the density estimation. \citet{bornkamp2009bayesiangermen} used a general discrete random measure introduced by \citet{ongaro2004discrete}. \citet{ohlssen2015flexible} introduced the assumption of Bayesian model averaging on  an experiment that has a control treatment as monotonicity dose-response model.

\subsubsection{AM-Spline} \label{Spline}

In our proposed spline we picked a fixed $\lambda$ and specific knots but the previous methods could be used with other application. AM-spline is our novel approach, using the normal CDF.  
The probability density function of the normal distribution with parameters $\mu$ and $\sigma^2$ is given by

${\displaystyle \operatorname {Normal} \sim (\mu ,\,\sigma ^{2})}$

\begin{equation} \label{eq:4}
f{(x)}=\frac{1}{\sqrt{2\pi\sigma x}} exp(-\frac{(x-\mu)^2}{2\sigma^2}),x>0 
\end{equation}

\begin{figure}[H]  
	\begin{center}
		\includegraphics[scale=0.4]{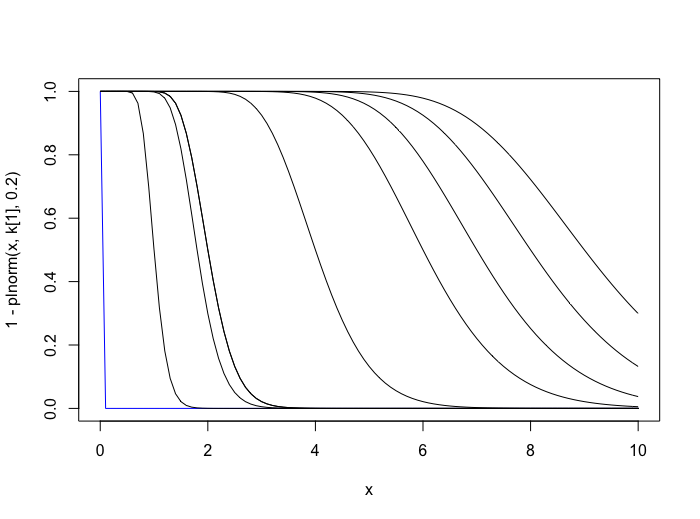}
		\caption{Spline Basis Functions }
		\label{fig: Spline Base Function}
	\end{center}
\end{figure}

And the base function for the AM-spline is  

\begin{equation} \label{eq:5}
CDF_{(normal)}={\displaystyle {\frac {1}{2}}\left[1+\operatorname {erf} \left({\frac {x-\mu }{\sigma {\sqrt {2}}}}\right)\right]} 
\end{equation}

where 
$$erf(x)=\frac{2}{\sqrt{\pi}}\int_{0}^{x}e^{-t^2}dt $$
The monotonic decreasing AM-spline is as the following  
\begin{equation} \label{eq:6}
f_{(norm)}=\sum a_{i} (1-F(x)), \quad \quad f(x)=(1-F(x))
\end{equation}
Where $a_{i}$ are the weights and $F(x)$ is the Normal CDF. The basis function could be any CDF of any statistical distribution. Figure \ref{fig: Spline Base Function} represent the basis functions of AM-spline each curve is one base function $f(x)$ in equation \ref{eq:6}. 

\begin{figure}[H]  
	\begin{center}
		\includegraphics[scale=0.4]{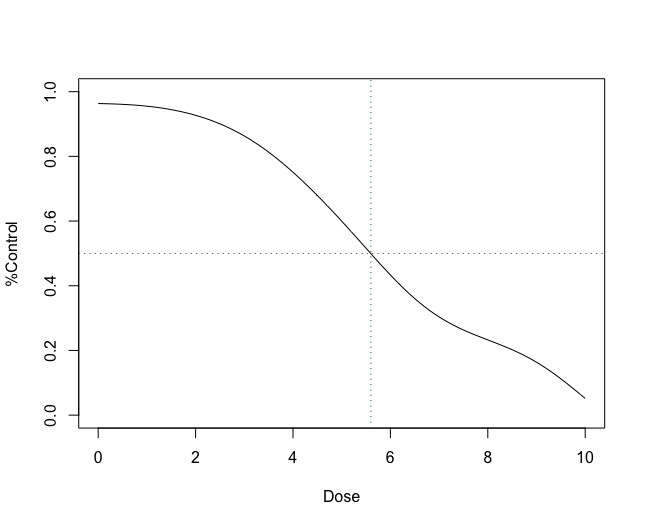}
		\caption{Fit of smooth monotonic decreasing function on the interval [0,1].  Two-dotted lines define the tolerable area under the fitted model as the following, The dotted horizontal line in the y-axis is the $ED_ {50} $ and the dotted vertical line in the x-axis is the dose correspond with the $ED_ {50} $ 
		}
		\label{fig: fit of NCDF}
	\end{center}
\end{figure}

General representation of the AM-spline in Figure \ref{fig: fit of NCDF}, which show the monotonic decreasing spline that fitted in the interval of $[0,1]$ as the solid curve and the dashed line represent the $ED_{50}$ under the fitted curve as if the spline used as a dose response model.  

\section{Bayesian Nonparametric} \label{Baysian}

The Bayesian framework is the road to get the full distribution of the tolerable region and quantify the uncertainty.   
\subsection{Prior Distribution}
Considering Bayesian approach since it updates the belief of a study in the guise of new data, where we could choose the prior and update the posterior distribution. We have our restriction in the AM-spline so, our distribution is in the interval of [0,1]. Dirichlet Process (DP) is a random probability distribution $F$ that generated by a PD on any partition $ A_{1}, A_{2}, ....A_{k}$ of the sample space which follows a Dirichlet distribution: $$(F(A_{1}),....,F(A_{K}) \sim D(a.F_{0}(A_{1}),....,a.F_{0}(A_{k})).$$ Where its defined as $F\sim \mathcal{D}(a,F_{0})$ and the parameters are the weight parameter $a$, and $F_{0}$ the base function. 
The DP is a stochastic process that used in Bayesian nonparametric models. It is the most popular Bayesian nonparametric models and sometimes its called the "distribution over distributions" since it can be thought as a distribution of probabilities themselves, and it has a variety of application and methods. \citet{zhou2018dirichlet} used the Dirichlet Process in mixture model with latent variables.

Dirichlet Process has several equivalent processes which could be a sampling methods such as: p\'olya urn, Chinese Restaurant Process, Hierarchical Dirichlet Process, Indian Buffet Process and the Stick Breaking process \cite{frigyik2010introduction}. In this paper, we are using the Stick-Breaking techniques as it is a representation of the DP. 
Our model has the Dirichlet distribution which is a generalization of the Beta distribution into multiple dimensions. Beta distribution ($\alpha, \beta$) is defined on (0,1) with density as 
$$f(x;\alpha,\beta)= \frac{\Gamma(\alpha+\beta)}{\Gamma(\alpha)\Gamma(\beta)}x^{\alpha-1}(1-x)^{\beta-1}$$
Note that if $X\sim Beta (a, b)$ then $\pi=(X,1-X)\sim Dir(\alpha)$ where $\alpha=[a,b]$ 

In other words, it is a distribution over the (K-1) dimensional simplex, this distribution over the value of K parameters who sum up to 1. It's parametrized by the K-dimensional vector $(\alpha_{1},\alpha_{2},...,\alpha_{K})$, where $\alpha_{K} \geq$ $0$ $\forall$ $K$ and $\sum_{K} \alpha_{K} > 0$. The distribution given by:

\begin{equation} \label{eq:7}
P(\pi_{K})= \frac{\prod_{K=1}^{K}\Gamma(\alpha_{0})} {\Gamma (\sum_{K=1}^{K} \alpha_{K})}\prod_{K=1}^{K} \pi_{K}^{\alpha_{k-1}}
\end{equation}
When $\pi \sim (\alpha_{1},.....\alpha_{K})$ then $\pi_{K}\geq \forall K $, $\alpha_{0}=\sum^{k}_{i=1} \alpha_{k}$ and $\sum_{K=1}^{K} \pi_{K}=1$. The expectation of this distribution is :
\begin{equation} \label{eq:8}
\mathbb{E}[(\pi_{1},...,\pi_{K})]= \frac{(\alpha_{1},...,\alpha_{K})}{\sum_{K}\alpha_{K}} 
\end{equation}

Dirichlet distribution is a conjugate prior for the multinomial distribution, more about Dirichlet Process prior in \citet{ferguson1973bayesian}.
If $\pi \sim Dir(\alpha_{1}, ..., \alpha_{K})$ and $x_{n} \sim Multi(\pi)$ are $ iid $ samples then is:

$$p(\pi | x_{1},.., x_{n})\propto p(x_{1},...,x_{n}|\pi)p(\pi)$$
$$= \bigg( \frac{\prod_{K=1}^{K}\Gamma(\alpha_{k})} {\Gamma (\sum_{K=1}^{K} \alpha_{K})}\prod_{K=1}^{K} \pi_{K}^{\alpha_{k-1}}\bigg)\bigg(\frac{n !}{m_{1}!...m_{K}!}\pi_{1}^{m_{1}}...\pi_{K}^{m_{K}} \bigg)$$
$$\propto \frac{\prod_{K=1}^{K}\Gamma(\alpha_{k}+m_{k})} {\Gamma (\sum_{K=1}^{K} \alpha_{K}+m_{k})}\prod_{K=1}^{K} \pi_{K}^{\alpha_{k}+m_{k-1}}$$
$$=Dirichlet(\alpha_{1}+m_{1},...,\alpha_{K}+m_{K})$$

Where $ m_{k}$ represent the counts of instances of $x_{n}=k$ in the data set.
Variational inference algorithm introduced by

Stick-breaking prior is a technique used to fit the posterior of Bayesian Nonparametric model using a Gibbs samplers tool. 
Stick-breaking prior is similar to the discrete random probability measures $C$, which has the general form
\begin{equation}\label{eq:1}
C(.)= \sum_{k=1}^{N} a_{k}\delta_{Q_{k}}(.),
\end{equation}
Where $ a_{k}$ is the random weight, $0 \leq a_{k} \leq 1$ and $\sum_{k=1}^{N} a_{k}=1 $. $\delta_{Q_{k}}(.)$ represent the discrete measure at ${Q_{k}}$ with $iid$ random assumption. $1 \leq N \leq \infty$ where $N$ could depend on the sample size $n$ in some cases and it could be finite or infinite. Equation \ref{eq:1} is similar to equation \ref{eq:3} but differ in the term of the multiplied term.
The random weight selection process is what distinguishes stick-breaking priories apart from the general random measures $C$ as in equation (\ref{eq:1}). If random probability measures or stick breaking random measures is in the form of equation (\ref{eq:1}) then call $C$ a $C_{N}(a,b)$ and 

\begin{equation}\label{eq:2}
C_{1}=Z_{1}\quad \textrm{and} \quad C_{k}= (1-Z_{1})(1-Z_{2}).......(1-Z_{k-1})Z_{k}, \quad \quad k\geq 2
\end{equation}
$Z_{k}$ $ \stackrel{ind}{\sim}$ $Beta(a_{k},b_{k})$, \quad $a_{k},b_{k} > 0 $ where $\mathbf{a}$ as vector $(a_{1},a_{2},..)$ and $\mathbf{b}$ as vector as well $(b_{1},b_{2},...)$. Stick breaking procedure could be in form of equation (\ref{eq:2}), the steps of stick breaking are considered independent and random properties when breaking what is left of a stick, and specifying the length of this break to the current $ C_{k}$ value. There is a long history of stick breaking weight determination mentioned in \citet{sta}. 
Two possible cases in stick breaking procedure

$N <\infty$ and $ N = \infty$ both use Dirichlet distribution on the random weight and have been introduced in \citet{sta}. More literature in stick-Breaking could be found in \citet{rodriguez2011nonparametric}, \citet{ishwaran2003some}. 

The posterior distribution is determined using the sampling method Metropolis Hasting (MH) which was introduced by \citet{hastings1970monte}. We used the Stick-breaking prior in our method which is a type of the MH sampling method. Nonparametric  model consider the Dirichlet distribution and its process which is a type of the Stick-Breaking process, as both used the beta-distribution random variables with the weight that drawn simultaneously by MCMC scheme, as \citet{kim2018nonparametric} and \citet{lynch2007introduction} used in determining the prior and the posterior.

\subsection{Posterior Computation }
Having the Dirichlet distribution as a prior distribution of the multinomial parameters $\pi \sim Dir(\alpha)$ and the posterior distribution is unknown. The parameters distribution are Dirichlet distribution which is different from those on the prior distribution. That makes posterior calculation straightforward, but in our study, it is difficult to find the posterior distribution. Therefore MCMC sampling methods are helpful in finding the predictive posterior distribution. The current observations are $X=\{x_{1},x_{2},....x_{N}\}$,\quad $\tilde{x}$ are the new observations (that the sampling method accept) and $\beta$ are the parameters.
\begin{equation} \label{eq:9}
p(\tilde{x})= \int_{\Theta} p(\tilde{x}|\beta,X)p(\beta|X)d_{\beta}
\end{equation}

Sampling algorithm:

\begin{algorithm}[H]
	\SetAlgoLined
	At each data points (parameter) :
	{We draw 10,000 MCMC samples
		\For{i=1,2,.... }{
			\begin{enumerate}
				
				\item For the posterior distribution with parameter $\beta$ drawn a sample $\beta^{m}$
				\item Stick-breaking is a sampling approach from Dirichlet process, which samples from thw distribution over the space and the distribution drawn is discrete with probability 1. The discrete distribution has a random probability mass function as
				$$f(\theta)=\sum^{\infty}_{k=1}\beta_{k}.\delta_{\theta_{k}}(\theta)$$
				Where $\beta_{k}$ are a random weights chosen to be independent of $\theta_{k}$ so as $0\leq \beta_{k} \leq 1$ and $\sum^{\infty}_{k=1} \beta_{k}=1$. Whereas, $\delta_{\theta k}$ is an indicator function which is zero everywhere, except $\delta_{\theta k}(\theta_{k})=1$

				$$\beta_{k}= \beta'_{k} \prod ^{k-1}_{i=1}(1-\beta'_{i}) $$
				Where $\beta'_{k}$ are independent random variables with the beta distribution $Beta(1,\alpha)$. $\beta_{k}$ is the length of the piece of a stick, process start with the stick unit-length and at each step stick break off a portion uniformly of the remaining stick $\beta'_{k}$ then assign to the broken-off piece to $\beta_{k}$. Accepting new candidate points if $0 < \beta' <1 $ 
				
				\item Posterior maintain  the Dirichlet process constraints        
				
			\end{enumerate}
		} 
		
	} 
	
	\label{olgorithm}
\end{algorithm}

More about both Gibbs sampling for stick-breaking prior given by \citet{sta} for MCMC methods by \citet{hastings1970monte}.

\subsection{BMD/BMDL estimation }
Dose response model has several famous threshold such as the Bench mark dose (BMD) and the No Observable Adverse Effect Level (NOAEL).
\citet{slob2005statistical} provides a comparison between BMD and NOAEL in multiple simulated studies. Bayesian BMD (BBMD) is a benchmark dose technique that incorporates the prior information. This can lead to saving animals lives in a study and improving the accuracy of the research \citet{slob2014shape}.

In this article the tolerable region is bounded by the $ED_{50}$ threshold, the region of the safest dose using the single chemical and the adverse effect. MCMC samples are  across all possible models, we use MH algorithm. The algorithm steps: the chain start with one set of mean $\mu_1$ in the simplest model and the variance drawn from the prior. Change is then introduced to the model, the changes are in adding a new set or rotating, change points or deleting existing set in the model. Accepted changes have probability $Q$  

\begin{equation} \label{eq:11}
Q= min\Big\{1,\frac{p(\beta'| Y) S(\beta| \beta')}{p(\beta| Y) S(\beta'|\beta)} \Big\} 
\end{equation}
Where $\beta$ is the model parameters current model and $\beta'$ is the new model with change parameters. $p(\beta| Y)$ is the posterior probability for the model parameters $\beta$ conditioning on the data. $S$ is the proposal distribution for the change parameters. When the changes are accepted then the chain move to the new $\beta'$ and if not the chain stay at it's study state $\beta$. These steps repeated for large number of times to get a samples $\beta_1 ,\beta_2,.....,\beta_i$. Success MCMC sampling method relays on the proposal distribution $S$, where it is effective in alternate the dimension samplers. Since adding and removing parameters from the model impact the likelihood of the new model \citet{holmes2003generalized}.

\section{Application }\label{Result} 
In this section, we will apply our proposed approach to evaluate its effectiveness on two simulated datasets and to the OP data introduced in Section \ref{ME}
\subsection{Simulated Example} \label{Simulated}
\begin{figure}[H] 
	\begin{center}
		\begin{tabular}{c c}
			(a) & (b) \\
			
			\includegraphics[width=3 in]{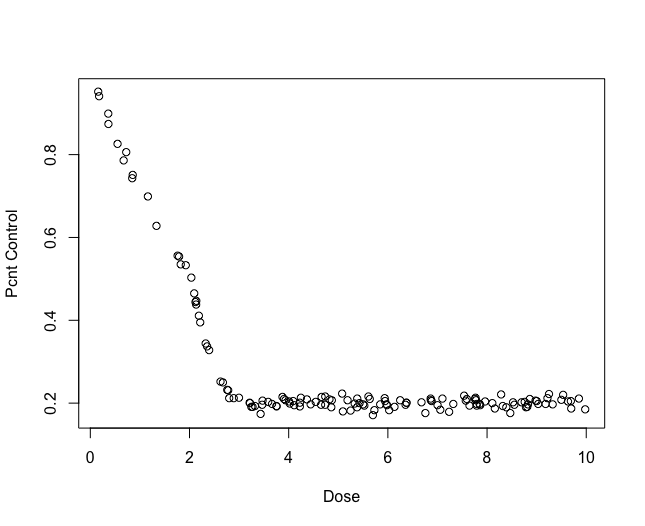}&
			\includegraphics[width=3 in]{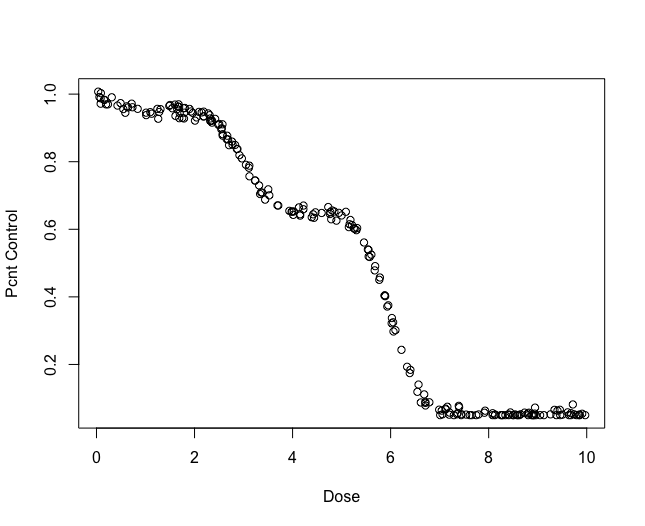}
		\end{tabular}
	\end{center} 
	\caption{Two simulated datasets Sim 1 (a) and  Sim 2 (b).}\label{fig:Simulated data}%
\end{figure}

To assess the viability of our proposed technique we applied it to two simulated datasets named \{sim 1\} and \{sim 2\}. These are pathological examples as they do not exhibit traditional parametric model shapes. Sim 1 has 140 observations and Sim 2 has 234 observations. Figure \ref{fig:Simulated data} shows the scatter plots of the two datasets; 
the x-axis is Dose, and the y-axis is the percent to control. 
In panel (a) Sim 1, the percent to control data decreases linearly as dose increases from  dosage $0$ to $3$. After dosage of $3$ the pattern levels out sharply to around $0.2$. In panel (b) Sim 2 has multiple inflection points at dosages $3$, $5$ and $7$. The percent to control is near zero from dosages $7$ to $ 10$. Notice that both of these simulated data are monotonically decreasing on the percent to control as the dose increases. 

\subsubsection{Sim 1 data analysis} \label{sssec:subsub1}

\begin{figure}[H] 
	\begin{center}
		\begin{tabular}{c c}
			(a) & (b) \\
			
			\includegraphics[width=3 in]{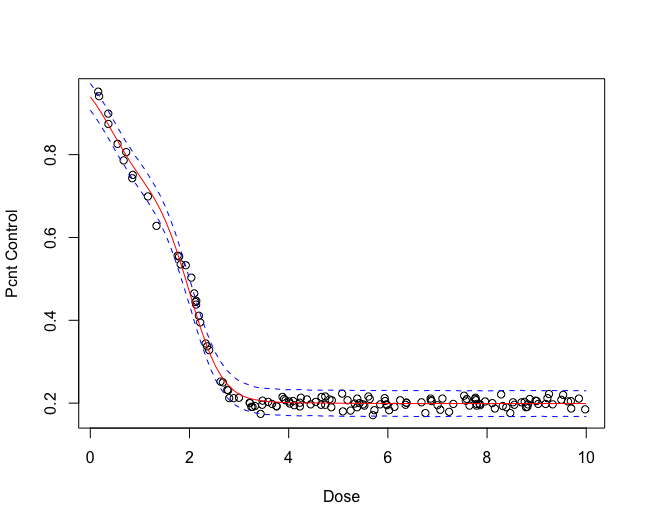}&
			\includegraphics[width=3 in]{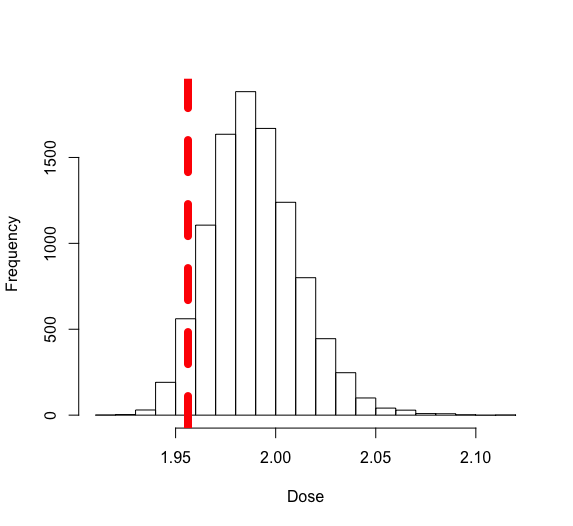}
		\end{tabular}
	\end{center} \caption{ Panel (a) AM-spline model fitted to the simulated Sim 1, the solid line represents the fit of the AM-spline and the dotted lines are the credible intervals. Panel (b) histogram represents the samples of the tolerable region and the vertical dashed line represents the $ED_{50}$.}\label{fig:data1 fit}
\end{figure}

To define the basis functions of the AM-spline model when applying it to the sim 1, we used the following knots k=\{ 0, 0.1, 0.25, 0.5, 1,1.2, 2, 3, 4, 5, 6, 7, 10\} and the bandwidth $\lambda= 0.8 $. We considered the normal likelihood and on the $a_{i}$ parameters, the stick-breaking prior distribution on the $\alpha$ parameters and chi-square distribution on the $\sigma$ parameters with $2$ degree of freedom. These priors combined with the likelihood are used to draw samples of the parameters from posterior distribution. The following parameters were estimated: the weights $a_{i}$ where $\sum_{i=1}^{k} a_{i}=1$, the standard deviation $\sigma $ and the lower support 
$\alpha$ using MCMC sampling methods. Python $3.6$ was used to implement the sampling scheme which took 30 minutes to obtain the samples. The Metropolis-Hastings sampling method was employed to obtain $10,000$ MCMC samples. Before running a long chain from the sampler we ran several short chains of $1,000$ to tune the sampler and then discarded these samples as burn-in samples. 
Table \ref{table:Data1} shows the estimate summaries of the posterior parameter samples of $\alpha$, $\sigma^2$ and the weights $a_{i=k}$. This shows that there are three parameters with heavyweights at the two knots $(k=\{0.5, 2\})$ and the $\alpha$. Fitting the AM-spline model to the simulated data sim 1 results in the parameters estimate represented in Table \ref{table:Data1} and the fit of the model in Figure \ref{fig:data1 fit}, 
which shows the AM-spline is fitted to Sim 1 data. In panel (a) the red solid line is the AM-spline mean, and the dashed lines are the $ 0.025$ and the $0.975$ quantiles of the posterior predictive distribution. We examined these convergence of the parameters of MCMC samples from the posterior distribution by visual inspection of the trace plots. These samples are used to generate samples from posterior predictive distribution which are then used to find the samples of the $ED_{50}$ distribution. $ED_{50}$ is the dosage with the $50\%$ corresponding response which define the tolerable region that calculated by Equation (\ref{eq:11}). Panel (b) represents the histogram of the tolerable region  
and the vertical dashed line on the histogram is the estimated $ED_{50}$. This is calculated as the $0.05$ quantile of the distribution effective dose which is at Dose= $1.95$. This demonstrates that the AM-spline is capable of finding an $ED_{50}$ for sim 1. 

Sim 1 is a fast decreasing simulated data that could exist in the toxicology lab. Determining a dose-response model for such data is difficult since no parametric or nonparametric  model has this type of function. AM-spline fits Sim 1 data adequately as shown in the Figure and table below.

\begin{table}[H]
	\caption{ Sim 1 Parameter estimates for the AM-Spline based on $10000$ MCMC samples from the posterior distribution, $\lambda = 0.8$, $\sigma=0.001$ and knots at $k=\{ 0, 0.1, 0.25, 0.5, 1,1.2, 2, 3, 4, 5, 6, 7, 10\}$} 
	\begin{center}
		\begin{tabular}{c c c c c c}
			\hline\hline
			Parameter  & Mean & Median &  StDev & 95\% Credible interval \\   \hline		
			$\alpha$ &  0.1989 & 0.1989 & 0.0017 & \textbf{( 0.1978, 0.2022)}\\
			$\sigma^2$ &  0.0002 & 0.0002 & 0.00003 & (0.0002, 0.0003)\\
			$a_{1}$ ($k=0$) &  0.0179 & 0.0142 & 0.0148 &  (0.0062, 0.0513)\\ 
			$a_{2}$ ($k=0.1$) &  0.0307 & 0.0246 &  0.0261 &  (0.0102,0.0930)\\ 
			$a_{3}$ ($k=0.25$) &  0.0541   & 0.0494 &  0.0372 &  (0.0238,0.1307 )\\ 
			$a_{4}$ ($k=0.5$) &  0.2287  & 0.2322 & 0.0434  &  \textbf{(0.1989,0.3032 )}\\ 	
			$a_{5}$ ($k=1$)  &  0.0157  & 0.0109 &  0.0148 &  (0.0046,0.0528 )\\ 	
			$a_{6}$ ($k=1.2$) &  0.0097  & 0.0067 & 0.0096  &  (0.0028,0.0354 )\\ 	
			$a_{7}$ ($k=2$) &  0.6166  & 0.6170 & 0.0198  &  \textbf{(0.6027,0.6523 )}\\ 	
			$a_{8}$ ($k=3$) &  0.0236  & 0.0242 & 0.0118  &  (0.0143,0.0444 )\\ 	
			$a_{9}$($k=4$)&0.0013&0.0008&0.0015&(0.0003,0.0052)\\ 					
			$a_{10}$ ($k=5$) &  0.0012  &  0.0008 &  0.0012 &  (0.0003,0.0042 )\\ 	
			$a_{11}$ ($k=6$) &  0.0003 & 0.00008 &  0.0005 &  (0.00001,0.0017 )\\ 	
			$a_{12}$($k=7$) &  0.0002 &0.00003 &0.0005 &(0.000008,0.0018)   \\  $a_{13}$ ($k=10$) & 0.00001  & 0.000005 & 0.00001 & (0.0000003,0.0001)\\ 
			
			\hline
		\end{tabular}
		\label{table:Data1}
	\end{center}
\end{table}

\subsubsection{Simulated Data 2 (Sim 2)}
In our second simulated data sim 2. We defined the basis functions of the AM-spline  by the following knots $k=\{ 0 ,3, 6\}$ and the bandwidth $\lambda=0.5$.
In sim 2 we used the exact same techniques as in sim 1, the only different was the type of data we have and the number of knots we used in sim 2. The data Sim 2 has more inflection points than in Sim 1, so that makes them different than each other.

\begin{figure}[H] 
	\begin{center}
		\begin{tabular}{c c}
			(a) & (b) \\	 
			\includegraphics[width=3 in]{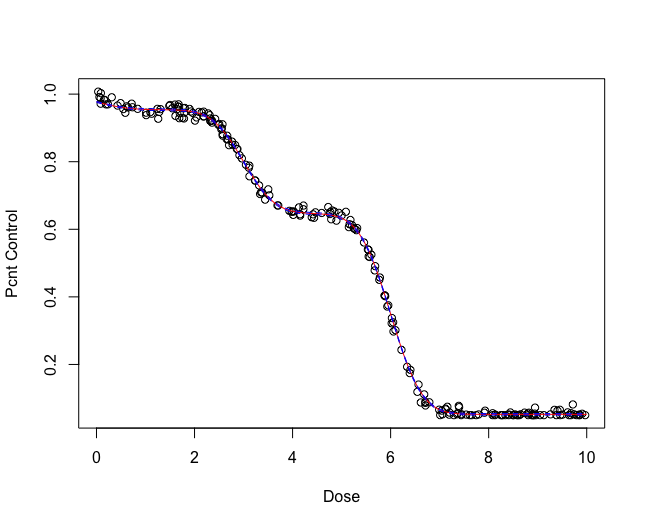}&
			\includegraphics[width=3 in]{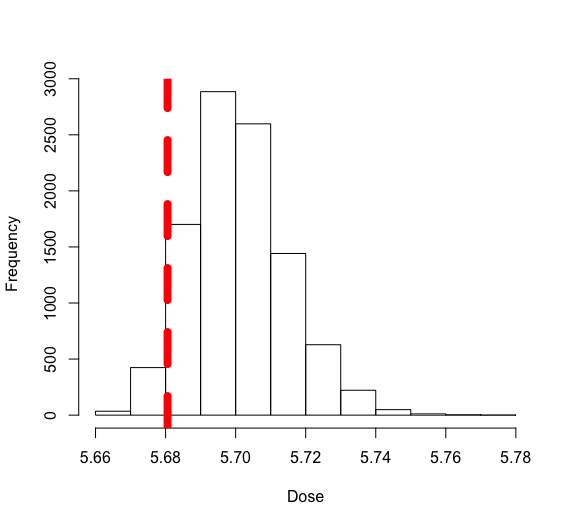}
		\end{tabular}
	\end{center} 
	\caption{Panel (a) AM-spline fitted model to the simulated Sim 2, the solid line represents the fit of the AM-spline and the dotted lines are the credible intervals and in panel (b) histogram represents the samples of the tolerable region and vertical dashed line represents the corresponding dose to the $ED_{50}$.}\label{fig:data2 fit}
\end{figure}

We considered the normal likelihood and the stick-breaking prior distribution on the $a_{i}$ parameters, the normal prior distribution on the $\alpha$ parameter, and chi-square distribution on the $\sigma $ parameters with 2 degree of freedom. These priors combined with the likelihood are used to draw samples of the parameters from the posterior distribution. 
The following parameters: $a_{i}$ are the weights where $\sum_{i=1}^{k} a_{i}=1$, $\sigma $ is the standard deviation and $\alpha$ is the lower support, were estimated using the MCMC sampling methods. Python $3.6$ was used to implement the sampling scheme which took 30 minutes to obtain the samples. The Metropolis-Hastings sampling method was employed to obtain $10,000$ MCMC samples. Before running a long chain from the sampler we ran several short chains of $1,000$ to tune the sampler and then discarded these samples as burn-in samples.  
Fitting the AM-spline model to the simulated data sim 2 results in the parameters estimate represented in Table \ref{table:Data2} and the fit of the model in Figure \ref{fig:data2 fit}. 
Table \ref{table:Data2} shows the summaries of MCMC parameter samples $\alpha$, $\sigma^2$ and $a_{i=k}$, which shows that AM-spline is fitted to Sim 2 data adequately and we see that two of the knots have heavyweights at $(k=\{3, 6\})$.   
Figure \ref{fig:data1 fit}
in panel (a) the red solid line is the AM-spline mean, and the dashed lines are the $ 0.025$ and the $0.975$ quantiles of the posterior predictive distribution. Since the fitted model is an adequate than the distribution of the tolerable region is calculated as in (\ref{sssec:subsub1}). We examined these convergence of the parameters of MCMC samples from the posterior predictive using the trace plots by visual inspection. These distribution samples were then of the tolerable region is defined by Equation (\ref{eq:11}) which determines the $ED_{50}$ dosage and the corresponding response across all MCMC samples. Panel (b) represent the histogram of the tolerable region,  
and the vertical dashed line on the histogram is the estimated $ED_{50}$; which calculated as the $5\%$ quantile of the distribution effective dose which is at Dose= $5.68$. This demonstrates that the AM-spline is capable of finding an $ED_{50}$ for sim 2.

Sim 2 is an example of a wiggle data that could exist in the toxicology lab. Determining a dose-response model for such data may be challenging,  AM-spline fits Sim 2 data adequately as shown in the Figure and table below. 

\begin{table}[H]
	\caption{ Sim 2 Parameter estimates for the AM-Spline based on $10000$ MCMC samples from the posterior distribution, $\lambda = 0.5$ and knots at $k=\{ 0 ,3, 6\}$} 
	\begin{center}
		\begin{tabular}{c c c c c c}
			\hline\hline
			Parameter & Mean & Median &  StDev & 95\% credible interval \\   \hline		
			$\alpha$ &  0.0523 & 0.0523 & 0.0011 & \textbf{( 0.0515, 0.0545)}\\
			$\sigma^2$  &  0.0001 & 0.0001 & 0.0001 & (0.0001, 0.0002)\\
			$a_1$ ($k=0$) &  0.0488 & 0.0488 & 0.0014 &  (0.0478, 0.0514)\\ 
			$a_2$ ($k=3$) &  0.3243 & 0.3243 &  0.0023 &  \textbf{(0.3228,0.3285 )}\\ 
			$a_3$ ($k=6$) &   0.6269 & 0.6269 & 0.0017  &  \textbf{(0.6258,0.6301 )}\\

			\hline
		\end{tabular}
		\label{table:Data2}
	\end{center}
\end{table}

From Figures \ref{fig:data1 fit} and \ref{fig:data2 fit}, we see that AM-Spline smoothly fit the data precisely where other model fail to fit the data as perfect as the AM-spline model, that is in the best of our knowledge. In Sim 2 the tolerable region is smaller than Sim 1 and that affected by the number of observation and numbers of knot. In sim 1 and sim 2 there are two knots that have a heavy weight.

\subsection{Organophosphate Data }\label{2.5}

The US EPA is interested in researching the effects of exposure to pesticides, especially Organophosphates (OP) such as Acephate (ACE) and Diazinon (DIA). \citet{moser2005neurotoxicological} conducted a laboratory study looking at Blood Cholinesterase as the endpoint when rats were dosed with these two chemicals more about the data in section \ref{ME}.
In this section, we Applied the univariate AM-spline model to the collected data. The goal here is to evaluate the effectiveness and the efficiency of AM-spline model and detriment the $ED_{50}$ in existing data.

\subsubsection{ ACE data}
\begin{figure}[H]  
	\begin{center}	
		\includegraphics[width=3 in]{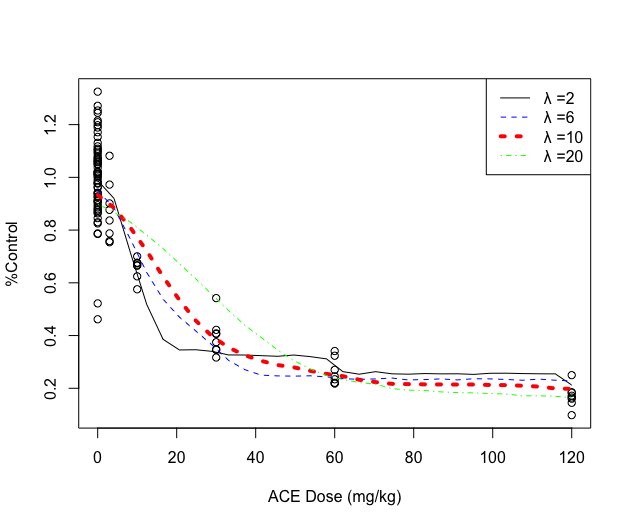}
		\caption{Comparing the fit of AM-spline model when considering the following four bandwidth \{$\lambda=(2,6,10,20 )$\}. We found $\lambda=10$ matched the data the most, so we used $\lambda=10$ as our bandwidth in this research}
		\label{fig: Compare model with different bandwidth }
	\end{center}
\end{figure}

The AM-spline model requires determining the smoothing parameter (bandwidth) $\lambda$, the set of knots and the stander deviation for each data. We fitted the following bandwidths:$\{2,6,10,20\}$ to determine the ideal bandwidth for the ACE data.
From figure \ref{fig: Compare model with different bandwidth } different bandwidth have been fitted to the ACE data, x-axis is the chemical doses and the y-axis is percent to control. The figure shows that when the bandwidth increases the smoothness of the model increases and when the bandwidth exceeds the data limit, the fitted model no longer fits the data well. Consequently, we picked $\lambda=10$ and The knots chosen randomly in the range of the data values, when fitting AM-spline model in this application. Since we are using the Metropolis Hastings algorithm and we use different coefficient step value when changing the bandwidth. 

\begin{figure}[H] 
	\begin{center}
		\begin{tabular}{c c}
			(a) & (b) \\
			\includegraphics[width=3in]{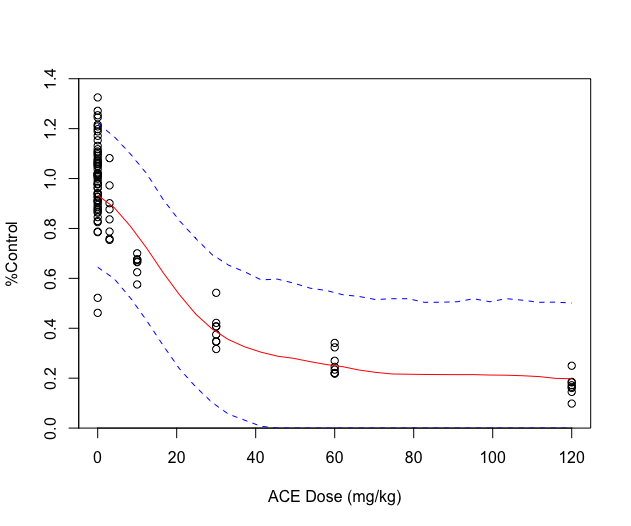} &
			\includegraphics[width=3in]{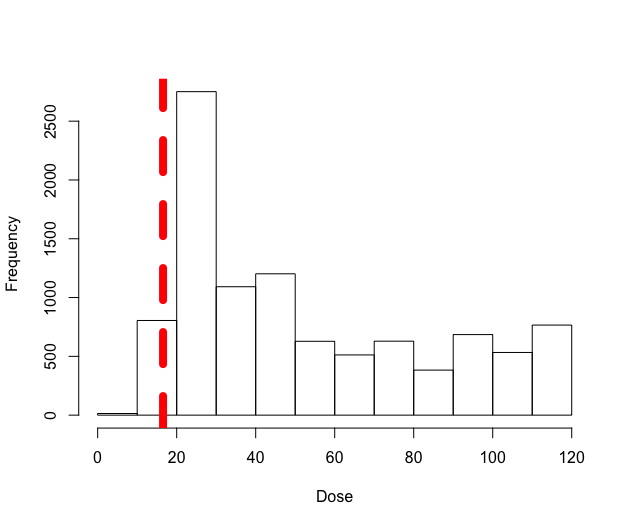}
		\end{tabular}
	\end{center} \caption{ Panel (a) AM-spline fitted to the ACE data as the solid line and the two dashed lines are the credible intervals and in panel (b) is the ACE-histogram of the samples of the tolerable region with the dashed line that represents the corresponding dose to the $ED_{50} $, the stressor (ACE) and the endpoint (BloodCHE). }\label{fig:ACE1}
\end{figure}

Applying the AM-spline to the ACE data using $\lambda=10$, and the knots as $k=\{0, 5, 9, 15, 30, 60 ,120\}$. The fit of the model will produce a parameters estimate and posterior predictive distribution. Therefore, we assumed normal likelihood, stick breaking prior in the weights parameters $a_{i}$, the normal prior on the lower support $\alpha$ and the chi-square prior in the stander deviation $\sigma$. To determine the posterior distribution we use the MCMC sampling tool to draw samples from the combined likelihood and priors. Using python $3.6$ which takes 45 minutes to obtain the samples. The Metropolis-Hastings sampling method used to obtain $10,000$ MCMC samples. Before running a long chain from the sampler, we ran several short chains of $1,000$ to tune the sampler and then discarded these samples as burn-in samples. Parameters estimate presented in table \ref{table:ACEM } which shows the estimate summaries of the posterior parameter samples of $\alpha$, $\sigma^2$ and the weights $a_{i=k}$. This table shows that there are three parameters with heavyweights at the knots $(k=\{15,30\})$ and the $\alpha$. and figure \ref{fig:ACE1} is the result of fitting the model to the data. Fitting the AM-spline model to the simulated data Sim 1 results in the parameters estimate represented in Table \ref{table:ACEM } and the fit of the model in Figure \ref{fig:ACE1}, which shows the AM-spline is fit to ACE 1. In panel (a) the red solid line is the AM-spline mean, and the dashed lines are the $ 0.025$ and the $0.975$ quantiles of the posterior predictive distribution. We examine these convergence of the parameters of MCMC samples from the posterior distribution by visual inspection of the trace plots.
These samples were then used to generate samples from posterior predictive distribution, which are used to find the samples of the $ED_{50}$ distribution. $ED_{50}$ is the dosage with the $50\%$ corresponding response, which define the tolerable region that is calculated by Equation (\ref{eq:11}). Panel (b) represents the histogram of the tolerable region, and the vertical dashed line on the histogram is the estimated $ED_{50}$. This is calculated as the $0.05$ quantile of the distribution effective dose which is at Dose = $18$. This demonstrates that the AM-spline is capable of finding an $ED_{50}$ for ACE data. 
\begin{table}[H]  
	\caption{ ACE data Parameter estimates for the AM-Spline fit considering  $10,000$ MCMC samples from the posterior distribution with $\lambda = 10$ and knots at $k=\{0, 5, 9, 15, 30, 60 ,120\}$} 
	\begin{center}
		\begin{tabular}{c c c c c c}
			\hline\hline
			Parameter & Mean & Median &  StDev & 95\% credible interval \\   \hline		
			$\alpha$ & 0.1776 & 0.1830 & 0.0576 & \textbf{( 0.1409, 0.2742)}\\
			$\sigma^2$ & 0.0212 & 0.0210 & 0.0030 & (0.0190, 0.0276)\\
			$a_1$ ($k=0$) & 0.0226 & 0.0180 & 0.0197 &  (0.0082, 0.0696)\\ 
			$a_2$ ($k=5$) & 0.0476 & 0.0350 & 0.0439  &  (0.0153, 0.1547)\\ 
			$a_3$ ($k=9$) & 0.1024 & 0.0819 &  0.0853 &  (0.0343, 0.3100)\\ 
			$a_4$ ($k=15$) & 0.5341 & 0.5487 & 0.1700  &  (0.4214, 0.8265)\\ 
			$a_5$ ($k=30$) & 0.1660 & 0.1374 & 0.1280  & \textbf{ (0.0618, 0.4560)}\\ 	        
			$a_6$ ($k=60$) & 0.0846 & 0.0692 & 0.0693  &  (0.0265, 0.2394)\\ 
			$a_7$ ($k=120$) & 0.0427 & 0.0216  & 0.0521  &  (0.0056, 0.1840)\\ 
			
			\hline
		\end{tabular}
		\label{table:ACEM  }
	\end{center}
\end{table}

\subsubsection{ DIA data} 
\begin{figure}[H]
	\begin{tabular}{c c}
		(a) & (b) \\
		\includegraphics[width=3in]{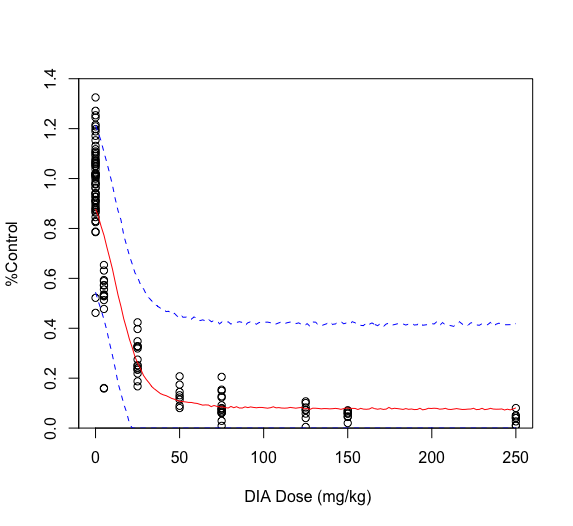} &
		\includegraphics[width=3in]{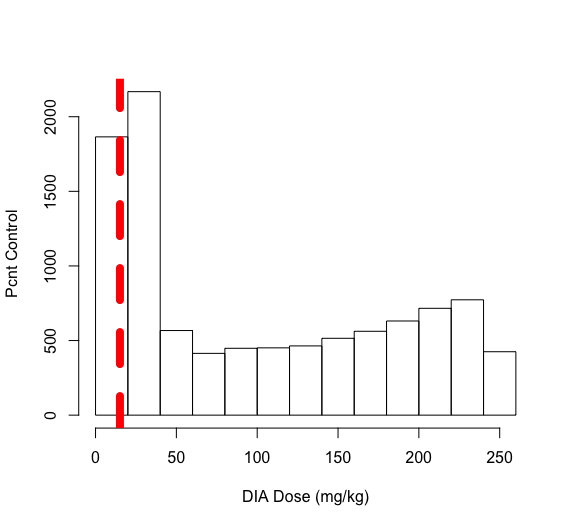} 
	\end{tabular}
	\caption{Panel (a) AM-spline fitted to the DIA data and in panel (b) histogram of $ED_{50}$, the stressor (DIA) and the endpoint (BloodCHE)}\label{fig:DIAFitHist}
\end{figure}

DIA is the second pesticide in OP data, DIA data is  second real data application to our model. Applying AM-spline to DIA data require using a specific smoothing parameter (bandwidth). We examine the following bandwidth :$\{2,6,10,20\}$ to determine the ideal bandwidth for the DIA data and we picked $\lambda=10$ since it fit the data the most. Am-spline fit the data considering $\lambda=10$  and the following knots $k=\{0, 5, 9, 15, 30, 60 ,120, 150, 250 \}$ which are in the data points. We have the following parameters that will be estimated: $a_{i}$ are the weights parameters that have the stick breaking prior. $\alpha$ the lower support parameter that has a normal prior and the stander deviation $\sigma$ parameter which has the chi-square prior with 2 degree of freedom. Combining the normal likelihood with the priors to draw samples using MCMC to define the predictive posterior distribution. 
The MH sampling algorithm used to obtain $10,000$ MCMC samples. Before running the full MCMC chain, we run several shorter chains of $1,000$ samples to tune the samples  and we discard them as a burn-in samples.
After fitting the model to the data using the provide information we get the following Table \ref{table:DIAM} and Figures \ref{fig:DIAFitHist}. Parameters estimate summaries of MCMC samples presented in Table \ref{table:DIAM} the $a_{i}=k$, $\alpha$ and $\sigma$. The table shows that two knots have heavy weights at ($k=\{9,15\}$). Figure \ref{fig:DIAFitHist} shows the fit of AM-spline model as in panel (a) the red solid line is the AM-spline mean, and the dashed lines are the $ 0.025$ and the $0.975$ quantiles of the posterior predictive distribution. We examine these convergence of the parameters of MCMC samples from the posterior distribution by visual inspection of the trace plots. These samples were then used to generate samples from posterior predictive distribution, which are used to find the samples of the $ED_{50}$ distribution. $ED_{50}$ is the dosage with the $50\%$ corresponding response, which define the tolerable region that is calculated by Equation (\ref{eq:11}). Panel (b) represents the histogram of the tolerable region, and the vertical dashed line on the histogram is the estimated $ED_{50}$. This is calculated as the $0.05$ quantile of the distribution effective dose which is at Dose = $20$. This demonstrates that the AM-spline is capable of finding an $ED_{50}$ for ACE data.

\begin{table}[H]
	\caption{ Parameter estimates for the AM-Spline considering $10,000$ MCMC samples from the posterior predictive distribution  using $\lambda = 10$ and the knots  $k=\{0, 5, 9, 15, 30, 60 ,120, 150, 250 \}. $}
	\begin{center}
		\begin{tabular}{c c c c c c}
			\hline\hline
			Parameter & Mean & Median &  StDev & 95\% credible interval \\   \hline		
			$\alpha$ & 0.0727 & 0.0736 & 0.0282 & ( 0.0538, 0.1232)\\
			$\sigma^2$ & 0.0292 & 0.0287 & 0.0038 & (0.0265, 0.0374)\\
			$a_1$ ($k=0$) &0.0316 & 0.0236 & 0.0275 &  (0.0101, 0.0986)\\ 
			$a_2$ ($k=5$) & 0.0821 & 0.0569 & 0.0723  &  (0.0240, 0.2415)\\ 
			$a_3$ ($k=9$) & 0.3654 & 0.3783 & 0.1696  &  \textbf{(0.2650, 0.6522)}\\ 
			$a_4$ ($k=15$) & 0.3591 & 0.3426 & 0.1712  &  \textbf{(0.2444, 0.7080)}\\ 
			$a_5$ ($k=30$) & 0.1182 & 0.1177 & 0.0544  &  (0.0769, 0.2257)\\ 	        
			$a_6$ ($k=60$) & 0.0331 & 0.0195 & 0.0372  &  (0.0072, 0.1338)\\ 
			$a_7$ ($k=120$) & 0.0039 &  0.0013 & 0.0052  &  (0.0002, 0.0170)\\ 
			$a_8$ ($k=150$) & 0.0018 & 0.0009 & 0.0024  &  (0.0002, 0.0087)\\ 
			$a_9$ ($k=250$) & 0.0048 & 0.0015 & 0.0065  &  (0.0001, 0.0201)\\ 	                
			\hline
		\end{tabular}
		\label{table:DIAM}
	\end{center}
\end{table}

From figure \ref{fig:ACE1} and \ref{fig:DIAFitHist} we see that the proposed model fit smoothly and perfectly to the two different data. 
In ACE the tolerable region is wider and high in magnitude compared to DIA and that affected by the range of the data.  

\subsection{Different ED's } 
From the above applications, we could calculate the following table of $ED_{90}$, $ED_{75}$ and $ED_{50}$ for each data individually. Since $ED_{90}$ and $ED_{75}$ are out of the dosage range, we found that $ED_{90}$ can not be determined or calculated for DIA data, data1 and data2.  So that tells us not all $ED's$ are possible to find, it depends on the available doses. In the other hand, $ED_{75}$ could be determined or calculated for data1 and data2, could be determined or calculated for data1 and data2, since the data has that possible doses. All ED's present in the following Table \ref{table:ED's all Data }.

\begin{table}[H]
	\caption{Different Effective Dose at each data set}\label{eval_table}
	\centering
	\begin{tabular}{|c|c|c|c|c|}
		\hline
		Data & ED &  Dose & 90\%CI \\
		\hline
		\multirow{2}{*}{ACE}&$ED_{50}$ &114.7147  & (90.451, 119.760)   \\
		\multirow{2}{*}{}& $ED_{75}$ & 39.7598 &  (18.258, 67.508) \\
		\multirow{2}{*}{}& $ED_{90}$ &17.5375  &(11.532,23.303)   \\
		\hline 
		\multirow{2}{*}{DIA}&$ED_{50}$ & 21.5215 &  (13.501, 191.016)   \\
		\multirow{2}{*}{}& $ED_{75}$ & 7.5075 &  (3.003, 10.010) \\
		\multirow{2}{*}{}& $ED_{90}$ & NA &  (, )\\
		\hline
		\hline 
		\multirow{2}{*}{data1}&$ED_{50}$ &5.9459  & (5.275, 6.006)   \\
		\multirow{2}{*}{}& $ED_{75}$ & N/A & (,)   \\
		\multirow{2}{*}{}& $ED_{90}$ & N/A & (,)  \\
		\hline
		\hline 
		\multirow{2}{*}{data2}&$ED_{50}$ &5.8559  &  (5.696, 6.116)  \\
		\multirow{2}{*}{}& $ED_{75}$ & N/A & (,)  \\
		\multirow{2}{*}{}& $ED_{90}$ & N/A &  (,) \\
		\hline
	\end{tabular}		\label{table:ED's all Data }
	
\end{table}

\subsection{Comparing AM-spline to Parametric methods }\label{compare}
\begin{figure}[H]
	\begin{center}	\includegraphics[width=3in]{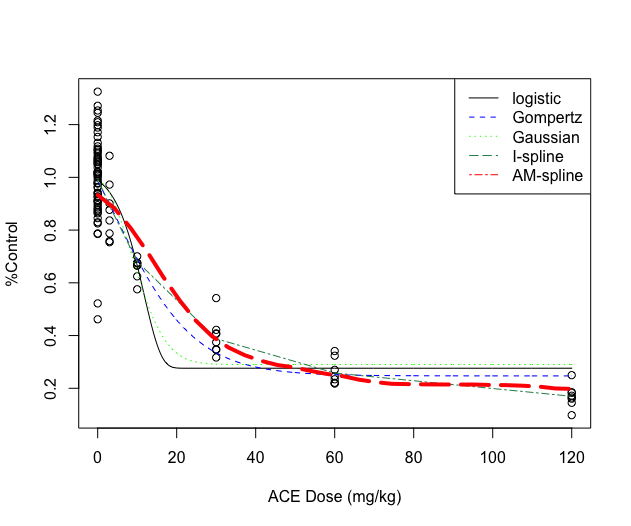}
		\caption{Gompertz, Logistic, Gaussian, Exponential, I-Spline and AM-Spline fitted to the OP data with ACE as the stressor and the endpoint BloodCHE as the response. }\label{fig:ACEdifferentDRmodel}
	\end{center} 
\end{figure}

\begin{figure}[H]
	\begin{tabular}{ c c }
		(a) & (b)  \\	
		\includegraphics[width=3.5in]{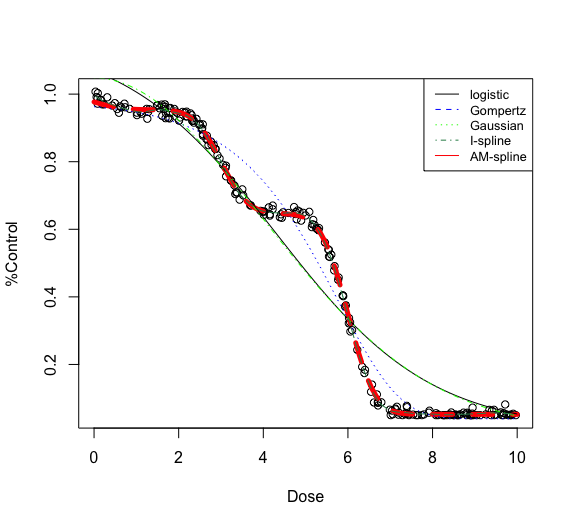} &
		\includegraphics[width=3in]{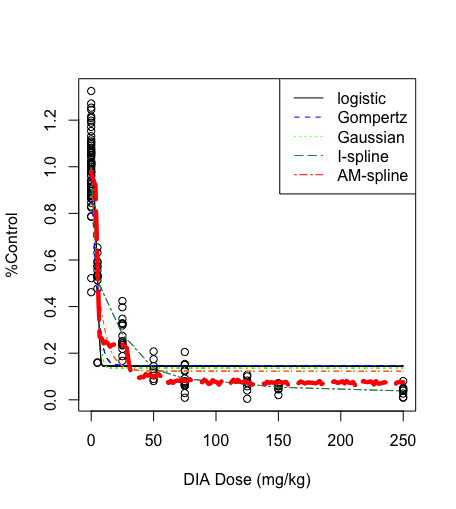} \\
		(c) & (d)  \\
		\includegraphics[width=3in]{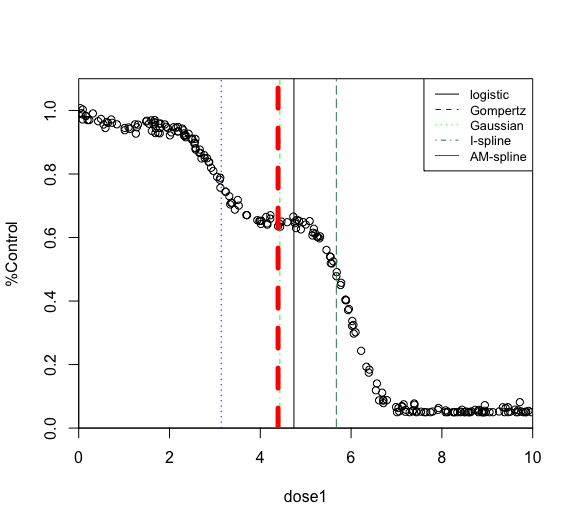} &
		\includegraphics[width=3in]{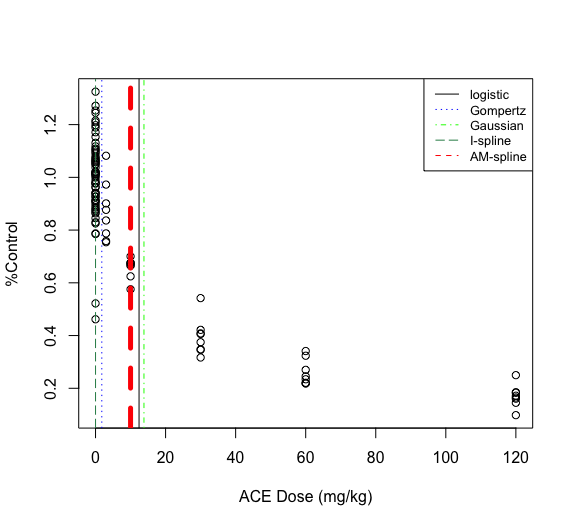}
	\end{tabular}
	\caption{$ED_{50}$ tolerable regions for Logistic, Gompertz, Gaussian, I-spline and AM-Spline for sim 2 data in panel (a) and panel(b) is the DIA data represent the endpoint BloodCHE. Vertical lines represent the $ED_{50}$ for each models in panel (c) and (d).}\label{fig:comp}
\end{figure}

Previous work has covered the parametric dose-response model Logistic, Gompertz, Gaussian. Our novel approach is the AM-spline for non-parametric dose-response model. Each model has a different representation so, which dose-response model performer better in fitting the most of the data? We used ACE data to answer this question, by fitting the different dose response models to the data. Figure \ref{fig:ACEdifferentDRmodel} shows the different fit of the models, x-axis the ACE dosage and the y-axis is percent to control. Our fitted model the AM-spline is the bold dashed line in figure \ref{fig:ACEdifferentDRmodel}, and it fit the data well. Whereas the other models performed considerably, but it did not fit the data well. For comparison, we randomly chose Sim 2 and the DIA data to compare the performance of the AM-spline along with other known parametric and nonparametric models. In Figure \ref{fig:comp} the different models are fitted to sim 2 as in panel (a) and to the DIA data as shown in panel (b), models are in different lines and colors. The bold dashed red line in the AM-spline model, it preform well in matching the data. Other models did not behave well and they were away from the data.  

in Figure \ref{fig:comp} we also compared the $ED_{50}$ for each data. The $ED_{50}$ for the chemicals using different models in different data 
From figure \ref{fig:comp} panel (c), we see the Gompertz model has $ED_{50}=3.3$ ,AM-spline and Gaussian models are performing at dose $4.5$. On the other hand, $ED_{50}$ for the Logistic model is at dose $4.9$ and the I-spline model is at dose $5.8$. In panel (d), the AM-spline is performing better than the Logistic and the Gaussian since it has a lower dose at $10$. Whereas, the Gompertz and the I-spline model are very close to the zero dose, which doesn't show the complete effect of the chemical. Therefore, AM-spline is minimizing the $ED_{50}$ dose and the adverse effect as well.

\section{Conclusions and Discussion  }\label{Discussion}

In this article, a new monotonic spline has been presented. We found that constrained smoothing splines are useful since we can customize the methods to any specific study and could choose the proper smoothing parameter $\lambda$.
Our nonparametric monotonically decreasing spline is flexible and effective in different statistical approaches and applications. Analysis and computation of nonparametric models are handier as mentioned in the literature The process and the fit of the model are sensitive to multiple properties: the step value, the size of the MCMC sample, the sampling starting value effect, the number of knots and the smoothing value. Our approach is producing new spline (AM-spline) with constraints on the distribution with values between zero and one. Fitting of the dose-response model to toxicology experiment was used to determine the tolerable region that defines the safest dose combination.  We limited the scope of this paper to the AM-spline introduction and there are more possible approach that could be developed in the future such as the penalized residual sum of square, experimental design approach, applied to different parameter selection method, spatially adaptive model and could be used as a model in the optimal follow-up design criteria.

%

%
%
%
%

\section{References}
\addcontentsline{toc}{section}{References}
\setcitestyle{numbers} 
\bibliographystyle{plainnat}

\bibliography{FirstREFprposal.bib}

\end{document}